# HELLENIC ARMS CONTROL CENTER

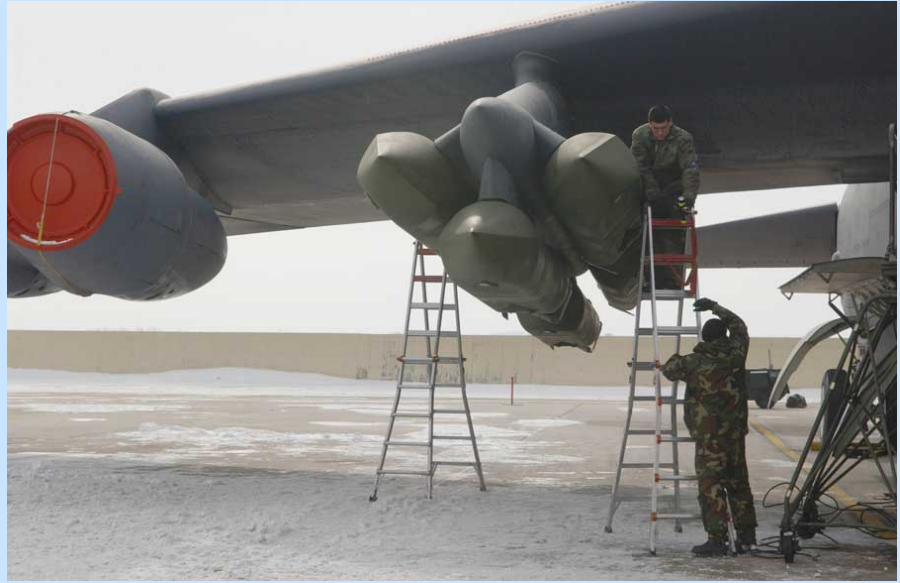

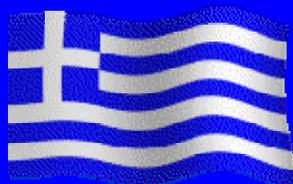

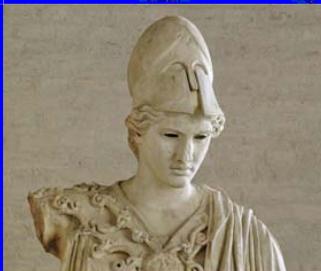

**Broken Arrows: Radiological hazards from nuclear warhead accidents (the Minot USAF base nuclear weapons incident)**

by Dr. Theodore E. Liolios

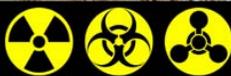

WWW.ARMSCONTROL.INFO

# HELLENIC ARMS CONTROL CENTER
# About ATHENA

*ATHENA*, **the Hellenic Arms Control Center,** is a non-profit, **nongovernmental**, scientific organization dedicated to promoting public understanding of and support for effective arms control policies. It is the only organization in Greece (and probably in the entire Europe) dedicated exclusively to research on arms control and nonproliferation issues. The Director of the Center is Dr. Theodore Liolios[1] currently a permanent member of the faculty of the Hellenic Army Academy and Director of its Nuclear Physics Laboratory.

*ATHENA* provides policy-makers, the press, and the interested public with authoritative information, analysis and commentary on scientific aspects of arms control, arms proposals, negotiations and agreements, and all relevant international security issues.

*ATHENA* holds regular press briefings on major arms control developments, providing commentary and analysis on a broad spectrum of issues for journalists and scholars both in Europe and abroad.

*ATHENA* covers numerous peace and security issues affected by proliferation of weapons of mass destruction including European and international nonproliferation programs, missile defenses, failed and post-conflict states and irresponsible defense spending.

The non-commercial academic nature of *ATHENA* ensures that all the proceeds from donations, subsidies and advertisements will be used to support the scientific research of the center (in the form of fellowships) as well as the education of young scientists (in the form of internships) who wish to specialize in arms control and international security subjects.

*ATHENA*'s homepage can also be considered a library, which provides information on various arms control topics. Most of that information has been adopted from some hot documents (which are readily downloadable from ATHENA'S homepage) and from many other internet scientific references

## *ATHENA*'s main objectives are:

▢ …to provide the international arms control community with a means of communicating their views online

▢ …to conduct scientific research in the field of arms control and non-proliferation

▢ …to improve the capabilities of the international intelligence community to respond to new and emerging threats, reducing the need to resort to the use of force, while enhancing the effectiveness of European military forces when needed.

▢ …to reduce the threat of weapons of mass destruction (WMD) and the risk of their use both by states who possess them and those ones seeking to acquire them.

▢ …to redirect the Hellenic military forces towards new capabilities aligned with the post-Cold War security environment, and to reduce the worldwide incidence of deadly conflict.

▢ …to provide citizens, decision-makers, scholars, and the press with accurate and timely information on nuclear, chemical, biological, and conventional weapons and strategies to reduce the dangers they pose.

▢ …to provide the entire world with a source of information on arms control issues.

For more information on the projects and publication of ATHENA contact:

ATHENA, HELLENIC ARMS CONTROL CENTER
THOMA XATZIKOY 11, 56122 THESSALONIKI, GREECE
TEL:+306944165341, FAX:+302310904794
ARMSCONTROL@ATH.FORTHNET.GR , WWW.ARMSCONTROL.INFO

---

[1] www.liolios.info

# HELLENIC ARMS CONTROL CENTER

## The Director's Message

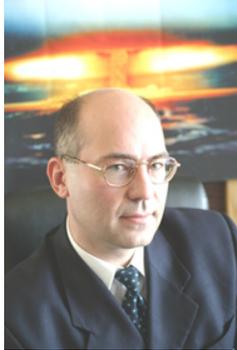

Nowadays, the most important issues that concern Europe and the entire world are terrorism, arms control and international security. Lack of education and information on the aforementioned subjects can lead to misinterpretation of intelligence which in turn can lead to the wrong political decisions.

The 9/11 terrorist attack introduced an era of uncertainty fear and awe for the civilized world. International terrorism has already attacked many European countries underlining the fact that counter-terrorism is not an American issue as is often suggested. Moreover, the proliferation of weapons of mass destruction (WMD) combined with the inadequate security measures in the countries which possess them enhances the possibility of a destruction of colossal dimensions. It is therefore obvious that Europe should actively embark on a campaign against WMD which will be precise and prudent where no collateral damage will be acceptable.

On the other hand, Europe, instead of passively following the doctrines of the USA, should emerge as an independent and powerful force in the new millennium. A unified well coordinated European Army should balance the universal influence of the USA. The European Armed Forces operating under the aegis of NATO, but without losing their autonomy and independence, could pose as a stabilizing and peace-keeping factor for the entire world.

Admittedly, the military excellence of the USA is largely due to the scientific excellence of their armed forces. It is therefore imperative that the European scientific community mobilizes in order to strengthen the European military science while at the same time informing the public and their governments about all arms control issues. This urgent necessity gave birth to *ATHENA*.

*ATHENA*, the Hellenic Arms Control Center, will attempt to challenge the American excellence in the field of Arms Control and Non-proliferation. Of course, this is not an easy task and it might be some time before *ATHENA* becomes part of the picture. In the mean time, we are confident that the scientific community, both in Europe and its allied states, will respond enthusiastically to our venture and become members of *ATHENA*.

*ATHENA* is a membership-based, non-profit, nongovernmental scientific organization which relies on your participation and contribution. All proceeds from contributions, donations, subsidies and advertisements will support the fellows, the interns and the activities of the Center. *ATHENA*'s non-commercial nature ensures that your contribution will be respected and used for a good cause.

Until *ATHENA* achieves its full potential there will be an incubation period, which calls for your lenience and patience.

Once *ATHENA* has become fully functional Europe will be proud of its first and most-ambitious Arms Control Center.

Join us and let's shape *ATHENA* together
.

With my Best Regards
 Dr. Theodore Liolios

# Broken Arrows: Radiological hazards from nuclear warhead accidents (the Minot USAF base nuclear weapons incident)

**Theodore E Liolios[1,2]**

[1]Hellenic Army Academy, Department of Physical Sciences & Applications, Laboratory of Nuclear & Atomic Phyiscs, Vari Attica 16673, Greece

[2]Hellenic Arms Control Center, Thoma Chatzikou 11, 56122 Thessaloniki, Greece

[1,2] URL:http://www.liolios.info, http://www.armscontrol.info

**Abstract.**
According to numerous press reports, in 2007 at Minot US Air Force Base six AGM-129 Advanced Cruise Missiles mistakenly armed with W80-1 thermonuclear warheads were loaded on a B-52H heavy bomber in place of six unarmed AGM-129 missiles that were awaiting transport to Barksdale US Air Force Base for disposal. The live nuclear missiles were not reported missing, and stood unsecured and unguarded while mounted to the aircraft for a period of 36 hours. The present work investigates the radiological hazards associated with a worst-case postulated accident that would disperse the nuclear material of the six warheads in large metropolitan cities. Using computer simulations approximate estimates are derived for the ensuing cancer mortality and land contamination after the accident. Health, decontamination and evacuation costs are also estimated in the framework of the linear risk model.

The analysis in the appendix of this study indicate that the six W80 nuclear warheads reportedly mounted on the six ACMs at the USAF Minot Base could not possibly contain more than 36 kg of Weapon-Grade Plutonium (WgPu) (six kg of WgPu each) with a more realistic approximate estimate of 18 kg WgPu (3kg of WgPu each).

Regarding inhalation hazards from such a postulated accident, even under the most unfavorable weather conditions (absolutely worst-case scenario), the simultaneous explosive dispersion of all the WgPu contained in the primary devices of the six thermonuclear warheads (maximum quantity of 36 kg WgPu) would generate a radioactive plume that could note pose any serious immediate threat (due to inhalation, submersion or ground shine) at distances larger than one kilometer from GZ. However, according to such a worst-case scenario, all the people at distances shorter than one kilometer from Ground Zero would be at a non-negligible risk of inhaling WgPu aerosols delivering a total dose which would increase the individual cancer risk by (at least) an amount of 5% -20%. The health costs per person at such distances could amount to (approximately) $1,500, or higher. All populated areas within a radius of one kilometer from GZ should be evacuated and the population should undergo medical tests to ensure that they have not inhaled any WgPu aerosols. In large metropolitan cities (1,000 people/sq.km) this amounts to 3,140 people and a total daily evacuation cost of $314,000.

However, regarding ground radiological contamination, realistic simulations indicate that there is non-zero probability that some areas within a radius of five kilometers from GZ can be so heavily contaminated with WgPu that their soil should be scraped, removed and buried in a safe location (e.g the **Savannah River Site**). Such a decontamination procedure, which should invariably be preceded by radiation detection procedures, would definitely force the authorities to evacuate the population within at least a radius of five kilometer from GZ.

Therefore, even if the WgPu inhalation hazard cannot extend to such large distances, ground contamination should definitely stretch evacuation distances to five kilometer from GZ. At distances larger than ten kilometers from GZ, there is a non-zero probability to find contaminated soil which should be watered and plowed and crops that should be removed and buried. No agricultural products from such areas should reach the market unless they have been screened for radiological contamination. Large scale emergency procedures at a metropolitan city (1,000 people per sq.km.) would entail the evacuation of citizens living in areas of seventy five square kilometers, that is 75,000 people. A daily total evacuation cost of $7,500,000 should be anticipated by the authorities while the decontamination operational costs

would be of the order of $75,000,000 (assuming arbitrarily an operational cost of one million per sq.km which should be scaled accordingly).

**Keywords:** nuclear weapons accidents, nuclear weapons, nuclear accidents,

1. Introduction

Almost immediately after the Thule "broken arrow" (US Department of Defense definition of a non-nuclear accidental detonation or burning of a nuclear weapon, DOD Directive 1993), the U.S. Air Force stopped routinely flying its bombers with nuclear weapons, and standard safety procedures were established which required that the warheads should have been removed from the missiles before they were attached to a bomber. According to numerous reports from the press (e.g. Starr 2007, Warrick and Pincus 2007 etc.) six W-80 (Mod 1) nuclear warheads were reportedly mounted (August 29-30, 2007) on six AGM-129 ACM cruise missiles and were mistakenly carried on a B-52H heavy bomber which flew from North Dakota to Louisiana. The USA government neither denied nor confirmed the incident which is its long standing policy. According to the press, the 2007 United States Air Force Bent Spear incident started at Minot Air Force Base in which six AGM-129 Advanced Cruise Missiles (ACM) mistakenly armed with W80-1[2] variable yield nuclear warheads were loaded on a B-52H[3] heavy bomber in place of six unarmed AGM-129 missiles that were awaiting transport to Barksdale Air Force Base for disposal. The live nuclear missiles were not reported missing, and stood unsecured and unguarded while mounted to the aircraft for a period of 36 hours.

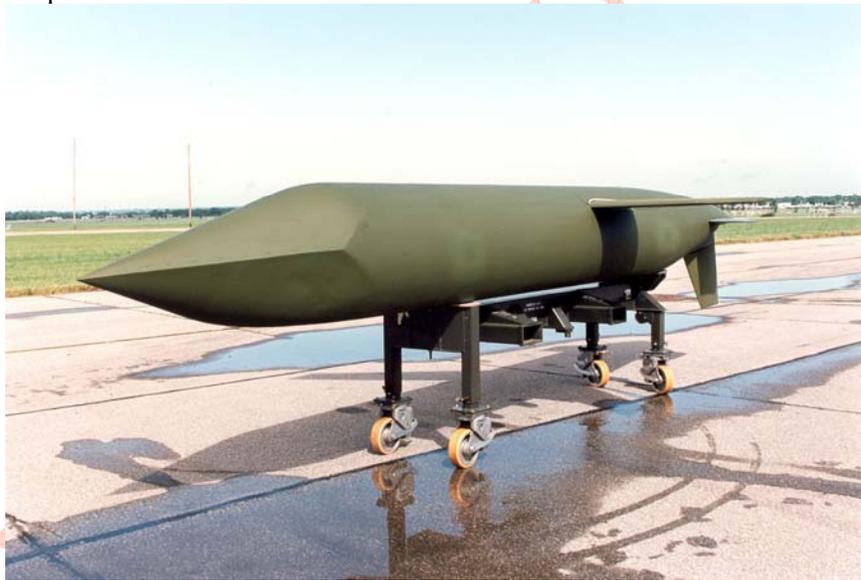

Picture 1. AGM-129A at the National Museum Of The AirForce (Wikipedia.org)

---

[2] The W80 is a small thermonuclear warhead (fusion weapon, or hydrogen bomb) in the nuclear stockpile of the USA with an adjustable explosive yield of between 5 and 150 kT TNT. The W80 is physically quite small, the "physics package" itself is about the size of a conventional Mk.81 250 lb (113 kg) bomb, 11.8 inches in diameter and 31.4 inches long, and only slightly heavier at about 290 lb (132 kg). Armorers have the ability to select the yield of the resulting explosion in-flight, a capability sometimes referred to as "dial-a-yield" but more properly Variable yield. At one end of the scale, perhaps using just the boosted fission primary, the W80 delivers about 5 kilotons of TNT, at the other it delivers about 150 kt. (wikipedia.org and references therein)

[3] Air Combat Command's B-52 is a long-range, heavy bomber that can perform a variety of missions. The bomber is capable of flying at high subsonic speeds at altitudes up to 50,000 feet (15,166.6 meters). It can carry nuclear or precision guided conventional ordnance with worldwide precision navigation capability (USAF B-52 Fact Sheet).

The Air Force's account of what really happened was provided by multiple sources who spoke to the press on the condition of anonymity because the government's investigation has been classified. Regardless of the details of this accident, it is now obvious that security measures in US nuclear weapon stewardship may have been breached and the public would like to know what hazards would really have been entailed if a nuclear warhead had been involved in an accident in this "Bent Spear" incident. There are numerous accident scenarios that might have actually turned this "bent spear"[4] incident into a "broken arrow"[5] one, such as mid-air collisions, accidental ejections of the missile, engine failure of the B-52 etc.

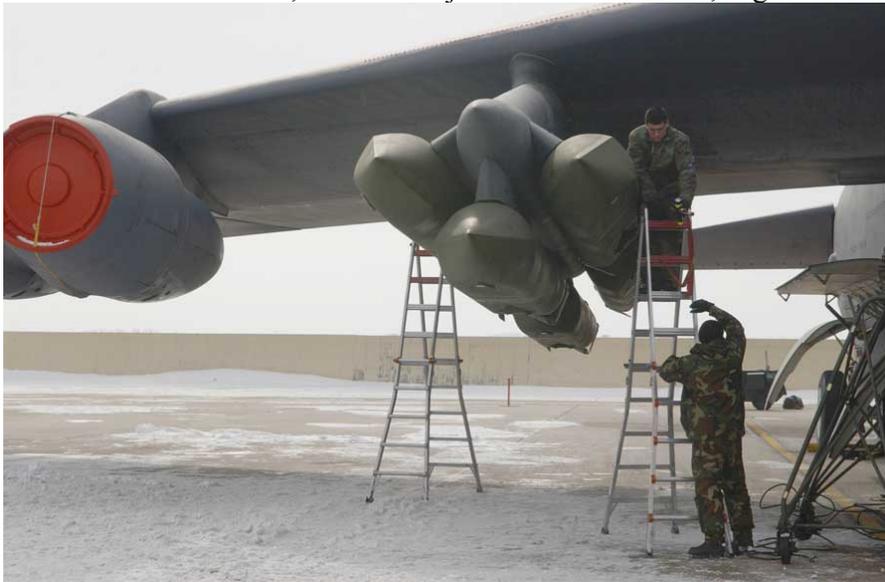

Picture 2. AGM-129A Cruise Missiles Being Secured on a B-52H bomber (Wikipedia.org)

In this study we will simulate a nuclear warhead accident and estimate the associated risks focusing in particular on the ensuing latent cancer mortality to human beings. Although nuclear weapon technology is classified, open literature abounds with relevant data which can yield very reliable estimates of the effects of an accident.
In addition to data from modern computer simulations, our study uses data collected from actual nuclear weapon accidents, and experiments (field tests) simulating nuclear warhead accident.

**2. Available data**

*2.1 Major Accidents*
Although there have been numerous nuclear warhead accidents in all nuclear states throughout history there are two major "broken arrow" incidents of which we are aware and in which the chemical high explosive (HE) in U.S. nuclear warheads exploded and contaminated an area with plutonium:
 a) In January 1966, over Palomares, Spain, a mid-air collision between a B-52 and its refueling aircraft resulted in four bombs from the B-52 being released. The braking parachutes of two bombs failed completely and they struck the ground at high speed spreading nuclear material in the vicinity of ground zero. Cleanup, weapon-retrieval and health operations cost $80 million while another $20 million dollars was spent in reparation and other costs. Of the four hydrogen bombs (Randall 1966) (designated during the recovery operations as weapon #1…#4) three were found on land, near the small fishing village of Palomares, part of Cuevas del Almanzora municipality, in Almeria, Andalusia (Spain). The fourth, which fell into the Mediterranean Sea, was recovered eighty days later. Weapon #2 experienced (Place W, Cobb F, and Deffending C 1975) an HE explosion case fragments and approximately 10 pounds of HE were found within 300 feet of its crater of about 20-foot diameter and 6 feet in depth. The tail section of #2 weapon had been displaced some 250 feet by the detonation. Weapon #3 also exploded on impact scattering approximately 80 pounds of HE and plastic within 100 feet of its crater. One fragment was found approximately 1500 feet from the crater. Radiation detection equipment detected significant alpha activity contamination in the area. About 4,600 separate 55 gallon drums of contaminated soil were shipped to Savannah River.

---

[4] "Bent Spear" refers to incidents involving nuclear weapons, warheads, components or vehicles transporting nuclear material that are of significant interest but are not of interest to the Major Commands, (Department of Defense and National Command Authority, United States Department of Defense directive 5230.16, Nuclear Accident and Incident Public Affairs Guidance)

[5] Broken Arrow refers to an accidental event that involves nuclear weapons, warheads or components, but which does not create the risk of nuclear war such as accidental or unexplained nuclear detonation., non-nuclear detonation or burning of a nuclear weapon, radioactive contamination etc..

b) In January 1968, near Thule, Greenland, a fire broke out on a B-52. The bomber was abandoned and crashed into the ice at high speed and burned; the HE in the four bombs it carried exploded, spreading plutonium over the ice (Project "Crested Ice" 1968). About 237,000 cubit feet of contaminated ice, snow and crash debris were shipped to Savannah River. Almost immediately after the Thule accident, the U.S. Air Force stopped routinely flying its bombers with nuclear weapons.

After the Palomares accident the Spanish and the US officials initially accepted the recommendation of the AEC (LANL) according to which all areas in which alpha counts per probe area are 100,000 cpm (770 $\mu g/m^2$)[6] or above will be removed to a depth of 5-6 cm and buried in an appropriate pit which will not permit seepage into the water table. All areas with counts between 100,000 cpm (770 $\mu g/m^2$) and 7,000 cpm (54 $\mu g/m^2$) will have the present crops removed and buried. All areas reading between 7,000 cpm (54 $\mu g/m^2$) and 500 cpm (3.85 $\mu g/m^2$) will be sprinkled with water to leach and fix the activity in the soil to minimize spreading by the wind. However, the Spanish desired much more conservative clean-up criteria than those recommended by Dr. W. Langham mainly for psychological reasons. According to the final agreement between the Spanish and the USA Government , the decontamination procedures that ensued were as follows:

| Table 1. Decontamination procedures adopted in the Palomares broken arrow event | |
|---|---|
| Soil Surface Contamination level | Decontamination Method |
| Above 462 $\mu g/m^2$ | Soil Scraped, removed and buried in a pit |
| Between 462 $\mu g/m^2$ and 5.4 $\mu g/m^2$ | Soil watered, plowed and monitored |
| Below 5.4 $\mu g/m^2$ | Soil watered and monitored |
| Below 77 $\mu g/m^2$ | Permissible when other methods not applicable |
| Vegetation with readings above 400 cpm (3.08 $\mu g/m^2$) | Vegetation removed and burned (or buried) |

2.2 Mitigation tests (Operation Roller Coaster)

Operation Roller Coaster was a field experiment at the Nevada Test Site (Shreve 1965, Stewart 1969). It involved the sampling and measurement of nuclear material dispersed by four separate test detonations. The detonations were triggered so as to simulate accidental detonation of the high explosives in the warheads. The experiment consisted of four separate detonations that simulated accidents involving different amounts of high explosives and different storage facilities (earthcovered structures, unbunkered buildings, open pads, or transportation vehicles).

The four tests were entitled Double Tracks, Clean Slate 1, Clean Slate 2, and Clean Slate 3. Double Tracks and Clean Slate 1 were unbunkered tests performed on open pads. Although there was some uptake of soil with these tests, fallout was not enhanced to the extent observed in the bunkered tests,

---

[6] Under perfect theoretical conditions of an infinite thin source of weapon grade uranium the correspondence would be 11,250 cpm and 100$\mu g/m2$. Any self shielding in the source or by ecological material would significantly change these figures. The correspondence for PAC-1S meters employed at Palomares is 100,000 cpm for a contamination of 770 $\mu g/m2$.

Clean Slate 2 and 3. Each test was conducted to simulate a storage or transport detonation accident. The entire experiment provided data over a range of cloud heights (a function of the amount of high explosives, Church H W 1969) and assessed the effect of plutonium-soil attachment (bunkered tests) on the fallout of particles near the detonation site. Declassified data from this operation are available and used in the Pantex Environmental Impact Statement (EIS) simulations. Other small-scale experiments are also cited in a study of plutonium fires (Condit 1993) which indicated that the fraction of plutonium converted into a respirable $PuO_2$ aerosol during a fire ranges from less than 0.001 percent to a few percent. However, when large quantities of explosives are involved in an explosive plutonium dispersion the amount of plutonium converted into $PuO_2$ aerosol is approximately (and can surely exceed) 20% of the total quantity involved. Therefore, it is now obvious why the present study doesn't focus on accidental plutonium fires

*2.3 The Pantex EIS simulations*

The Pantex Plant is a nuclear weapons assembly/disassembly facility located 25 km to the east-northeast of Amarillo, Texas. A series of reports was published by the staff of the Pantex Plant (Dewart *et al* 1982, Elder *et al* 1982, McDonell and Dewart 1982, Wenzel 1982) in support of preparation of an Environmental Impact Statement (EIS) required by the US government.

These reports, inter alia, cover the calculation of atmospheric dispersion and deposition of plutonium following postulated nonnuclear detonations of nuclear weapons (Dewart *et al* 1982). The computer code used to perform these calculations is the DIFOUT model (Luna 1969), developed at Sandia National Laboratories in conjunction with Operation Roller Coaster. Actually, DIFOUT is a tilting plume Gaussian dispersion model (Hanna *et al* 1982), which includes aerosol depletion through particle fallout and the effects on dispersion of wind speed and direction variation with height. The aerosol cloud produced by the detonation is divided into several horizontal cylindrical layers, each containing a specified amount of the total aerosol of the cloud. The aerosol is dispersed from a vertical line source in each layer; the downwind integrated air concentrations and ground deposition are a sum of the contributions from the line source in each layer. The calculated values of total airborne dosage and deposition dosage adequately approximated or conservatively overestimated (Dewart *et al* 1982) the measured values of Operation Roller Coaster.

**3. Simulating "Broken Arrows"**
A "broken arrow" in nuclear weapons terminology is practically a nuclear accident which can also be classified as a "dirty bomb" involving the (usually explosive) dispersion of weapon-grade plutonium (WgPu) and other materials, radioactive and inert, which could be dispersed along with the WgPu by the detonation accident. Apart from WgPu, some nuclear warheads may also contain other toxic materials (Elder *et al* 1982, Fetter 1990) such as uranium, tritium[7], beryllium as well as small amounts of fission products (notably iodine, strontium, and cesium) from a fission yield not exceeding $2.5 \times 10^{17}$ fissions per warhead (one-point-safe technology, see appendix). Regarding "broken arrows" the hazards associated with human exposure to these products are negligible compared to the WgPu radiological ones (Elder *et al* 1982) and will not be considered in this study.

Since the invention of nuclear weapons many theoretical models have been devised which can predict and estimate the risk associated with a nuclear weapon accident. In our study we use the results of two popular computer codes which have both been used by the US Department of Defense[8] namely

---

[7] In a typical modern boosted fission primary, a few grams of deuterium-trimium gas are injected into the center of a hallow core of plutonium immediately prior to the detonation of the surrounding chemical explosives.

[8] "Nuclear Weapon Accident Response Procedures (NARP)", DoD 3150.8-M, 1999

HOTSPOT[9], DIFOUT (Dewart *et al* 1982, Luna 1969, Elder *et a* 1982) as well as results from a naive model (cylinder model) which, despite its simplicity, can yield a fairly satisfactory picture for radiological hazards at close distances from Ground Zero (GZ)

3.1 Definition of a worst-case "broken arrow" scenario

In "broken arrow" radiological accident simulations the decisive parameters that will play a vital role in the subsequent health consequences and the degree of contamination are:
- ❖ the type and the quantity of the nuclear material in the warhead(s)
- ❖ the energy of the explosion that will disperse the nuclear material
- ❖ the meteorological conditions at the time of the accident
- ❖ the population density in the vicinity of the accident
- ❖ the degree of warning the population will receive before during and after the accident
- ❖ the quality of the hazard prepardness and the protective actions taken before, during and after the accident.

In a hypothetical worst-case "broken arrow" that might have ensued from the recent Minot Base "Bent spear" event:
1) All the ACM would be ejected from the bomber and strike the ground at high speed causing the detonation of the high-explosives in the warheads[10]. The impact would cause a chemical explosion in the six bomb(s) (originating from the high explosives in the warheads and the kinetic energy of the impact), spreading weapon-grade plutonium[11] widely around the point of impact.
2) all the weapon-grade plutonium in the six nuclear warheads would be explosively dispersed in the vicinity[12] with no warning[13] to the population in the vicinity of ground zero (GZ).
3) the explosion would be such that a great amount of the WgPu of the warheads would be rendered airborne and respirable[14]. The WgPu in the six warheads can be explosively dispersed due to a confined single explosion (e.g. Thule Greenland "broken arrow") or it can be dispersed during six different chemical explosions at an equal number of impact points (e.g. Palomares "broken arrow").
4) the accident would happen in a densely populated area during the daytime so that the population density in the vicinity of the accident would be maximum (e.g. during the rush hour).
5) the aerosols formed would be so fine that the plume would travel very far away from ground-zero without being depleted considerably

---

[9] https://www-gs.llnl.gov/hotspot/index.htm

[10] In an equally worse-case scenario the B52H bomber would be abandoned and crash into the ground at high speed

[11] Other nuclear material would also contaminate the area around GZ such as uranium isotopes and a few grams of tritium. The radiological hazard presented by such isotopes is much smaller than that of weapon-grade plutonium and will not be discussed in this study.

[12] Approximately 3 kg of WgPu per W80 Mod-1 warhead would amount to the dispersion of 20 kg of WgPu (see appendix)

[13] If the public has early enough warning it can simply evacuate the area thus avoiding any exposure to the lethal effects of accident. Even on a very short notice the public can simply resort to shelters or stay indoors during the plume passage. The radioactive plume will simply pass over the city and after some hours the air will be much safer. As regards warning we will assume that the public is totally unaware of the accident throughout the plume passage.

[14] Based on experiments and calculations, it has been estimated that 10–100 percent of plutonium contained in the warheads, with a best estimate of 20 percent, could be converted by such explosions into a PuO2 aerosol of respirable size (median aerodynamic diameter in the range of 5 microns or less) (***Supplementary Documentation for an Environmental Impact Statement Regarding the Pantex Plant,*** ibid)

6) in the vicinity of the accident there would be a very calm wind[15] which would slowly carry the plume downwind
7) there would be an atmospheric stability category which would maximize air concentrations and the subsequent doses of course near the receptors height.
8) there would be a narrow inversion layer[16] which would trap the WgPu aerosols and greatly increase its concentration between the ground and the top of layer.
9) the energy of the explosion[17] will be just about enough to aerosolize the entire quantity of WgPu. The explosion energy, due to the large uncertainties associated with it could be considered a free papameter. However, in a worst-case scenario it is reasonable to assume a very small energy yield[18] since experiments have shown[19] that the larger the explosive energy yield the smaller the concentrations and the relevant doses resulting from the accident[20].

## 4. HOTSPOT vs DIFOUT simulations

The Hotspot Health Physics codes were created to provide emergency response personnel and emergency planners with a fast, field-portable set of software tools for evaluating incidents involving radioactive material. The software is also used for safety-analysis of facilities handling nuclear material. Hotspot codes are a first-order approximation of the radiation effects associated with the atmospheric release of radioactive materials. Hotspot is a hybrid of the well-established Gaussian plume model, widely used for initial emergency assessment or safety-analysis planning. Virtual source terms are used to model the initial atmospheric distribution of source material following an explosion, fire, resuspension, or user-input geometry.

According to the HotSpot notes[21], for deposition velocities less than 0.1cm/sec and release points at or very near the ground level the maximum air concentration and ground surface deposition is always associated with F stability. Moreover, stability class F is usually accompanied by inversions. However, unlike stability categories A to D, stability class F rarely occurs during daytime therefore as regards the worst-case time for such a "broken arrow" there is a competition between the stability classes: Daytimes (i.e. Stability categories A, B) are associated with an increased population density in the streets and relatively low concentrations (and doses) while nighttimes (i.e. stability categories D, F) are associated with a decreased population density but relatively high concentrations (and doses). We have plotted isodose contours for all stability categories A-F under the unfavorable "broken arrow" scenario described in figures Fig.1-Fig.4 and have found that for the same population densities stability categories A, B result in collective doses which can be five times smaller than those under stability categories D and E. In this study stability category F is adopted in all worst-case HOTSPOT scenaria (represented by the solid curves in Figs.1-8).

---

[15] Wind speed is one of the most decisive parameter of a nuclear accdient. In fact if we assume that: (a) all other meteorological parameters are constant and (b) the radiological material has a relatively large half-life then the dose received at a certain distance downwind from GZ is inversely proportional to wind speed while the area receiving a certain dose is also a rapidly decreasing function of speed. Large wind speeds will quickly disperse the material at large distances thus lowering the average air concentration of the toxic substances while low wind speeds will have the opposite effect.

[16] We assume a worst-case mixing height of 300 m.

[17] In the Palomares "broken arrow" event the high explosives found in the vicinity of the weapon #2 crater were approximately 10 pounds, while the high explosives found in the vicinity of weapon #3 crater were approximately 80 pounds (including plastics). Therefore, as a plausible first approximation we can assume that the each warhead carried a quantity of at least 80 pounds of high and plastic explosives. An equally reasonable assumption would be that the WgPu in weapon #2 was dispersed with an explosion of at least 70 pounds of HE. High explosives are generally more energetic than TNT explosives (see for example J.Petes 1986), thus it is reasonable to assume that during a "broken arrow" incident of a single nuclear warhead an explosion of (at least) 70 pounds TNT can possibly occur. (Fetter S and Hippel F V 1990) have also estimated a plausible minimum energy yield of 88 pounds TNT equivalent.

[18] Each of the W80 Mod-1 warheads involved in the current postulated broken arrow incident reportedly weigh 132kg (wikipedia.org) and were mounted on AGM-129 ACM cruise missiles which according to a relevant USAF Fact Sheet (www.af.mil) weighs at least 1500 kg. Ejecting such an ACM from an altitude of 1000 m and ignoring air resistance the impact of the missile on the ground would cause an explosion of 3.5 kg TNT. The explosive yield scales linearly with altitude thus ten times higher altitudes would cause ten times more energetic explosions. Adding the explosive energy that may be released from the fuel and the high explosives in the primary device of the warhead we can obviously assume that an explosive dispersion of WgPu can easiliy occur at energies much larger than 10 kg TNT.

[19] *Supplementary Documentation for an Environmental Impact Statement Regarding the Pantex Plant,* Los Alamos National Laboratory, report LA-9445-PNT-D, 1982

[20] This is reasonable as the larger the TNT energy the larger the initial radioactive cloud which means that the WgPu will be distributed in a larger volume.

[21] https://www-gs.llnl.gov/hotspot/

As we have already mentioned, one of the most reliable studies of nuclear warhead accidents is a series of reports (Dewart *et al* 1982, Elder *et al* 1982, McDonell and Dewart 1982, Wenzel 1982) documenting work performed in support of preparation of an Environmental Impact Statement (EIS) regarding the Department of Energy (DOE) Pantex Plant near Amarillo, Texas. In particular, the report covering the calculation of atmospheric dispersion and deposition of plutonium following postulated nonnuclear detonations of nuclear weapons is very relevant to our simulations. In the Pantex Plant report, downwind total integrated air concentrations and ground deposition values for each postulated accident are derived using the DIFOUT model, developed at Sandia National Laboratories in conjunction with Operation Roller Coaster (Shreve J D 1965), a field experiment involving sampling and measurements of nuclear material dispersed by four detonations. In the same report, the DIFOUT model is described along with the detonation cloud sizes, aerosol parameters, and meteorological data used as input data. Finally, a

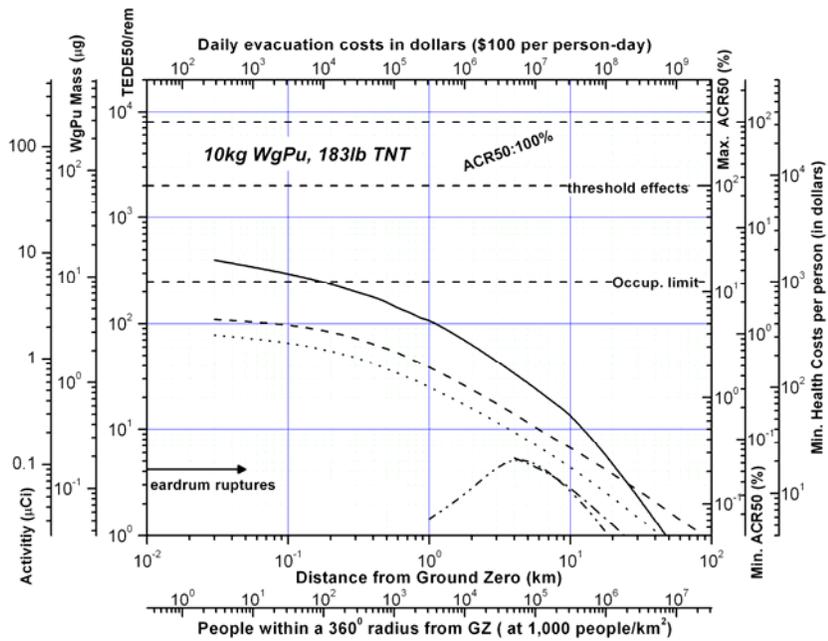

Fig.1. TEDE50 with respect to distance from GZ (183 lb TNT)

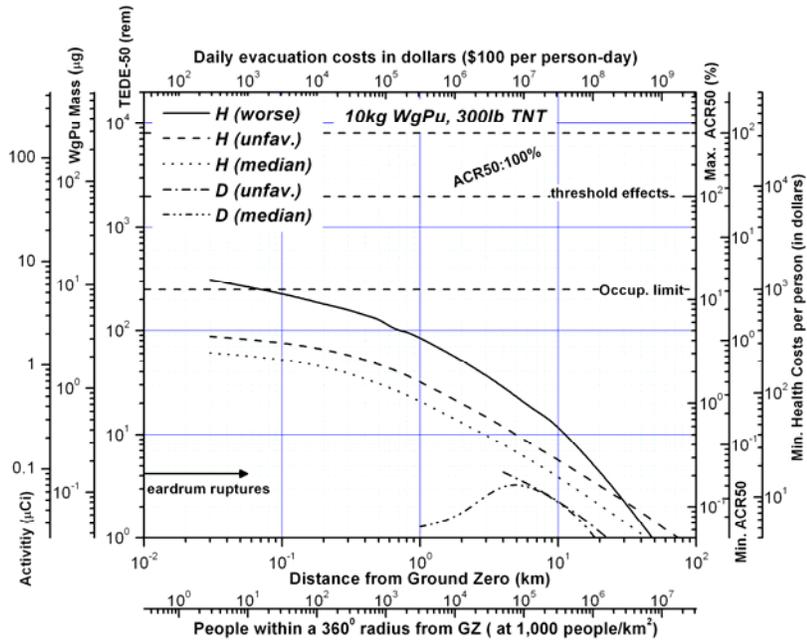

Fig.2. TEDE50 with respect to distance from GZ (300 lb TNT)

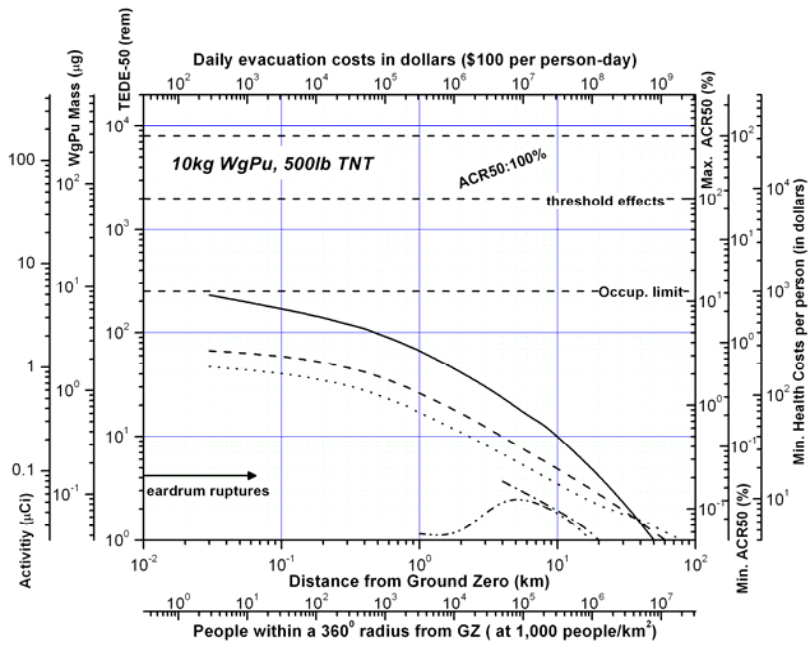

Fig.3. TEDE50 with respect to distance from GZ (500 lb TNT)

Page 13 of 30

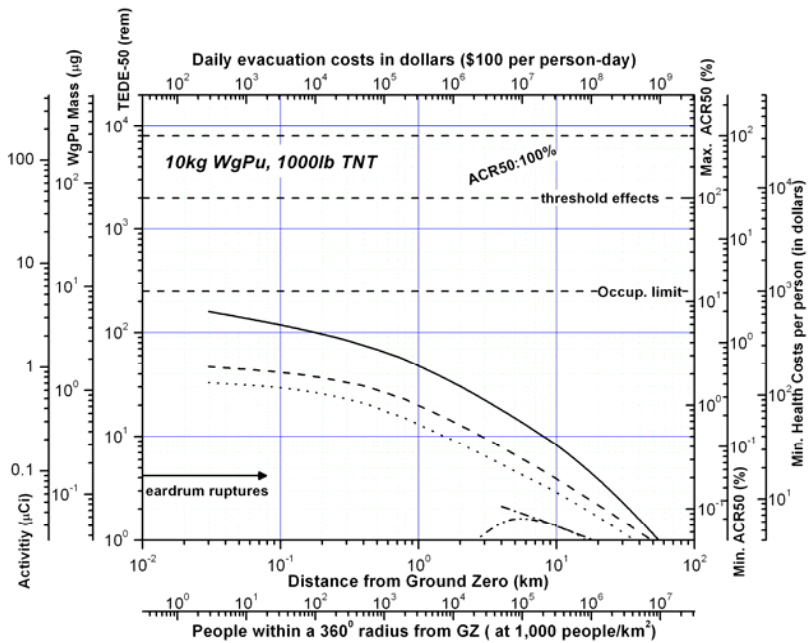

Fig.4. TEDE50 with respect to distance from GZ (1000 lb TNT)

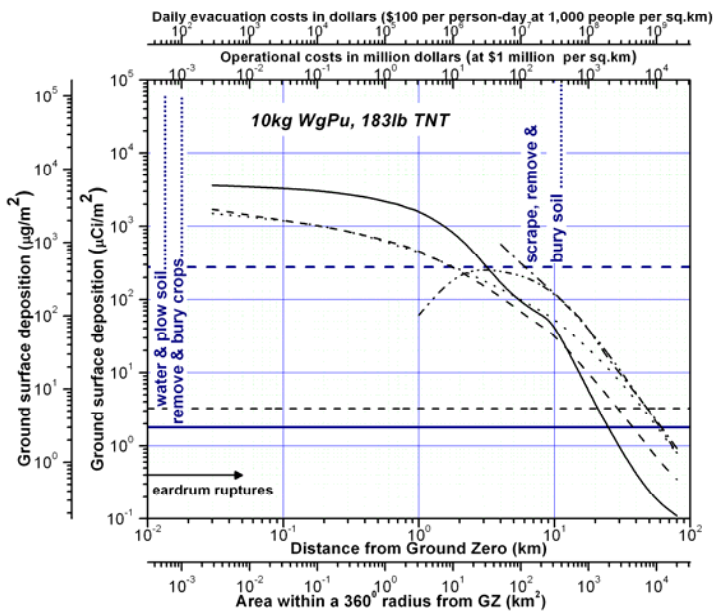

Fig.5. Ground surface deposition with respect to distance from GZ (183 lb TNT)

Page 14 of 30

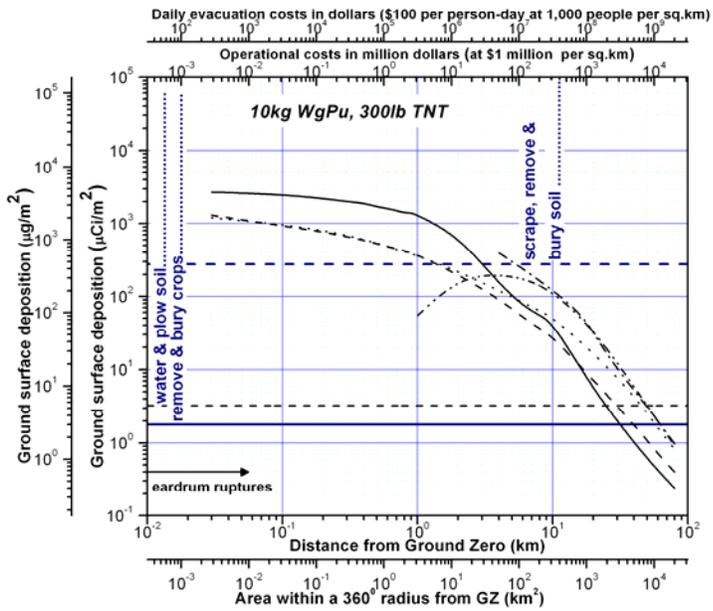

Fig.6. Ground surface deposition with respect to distance from GZ (300 lb TNT)

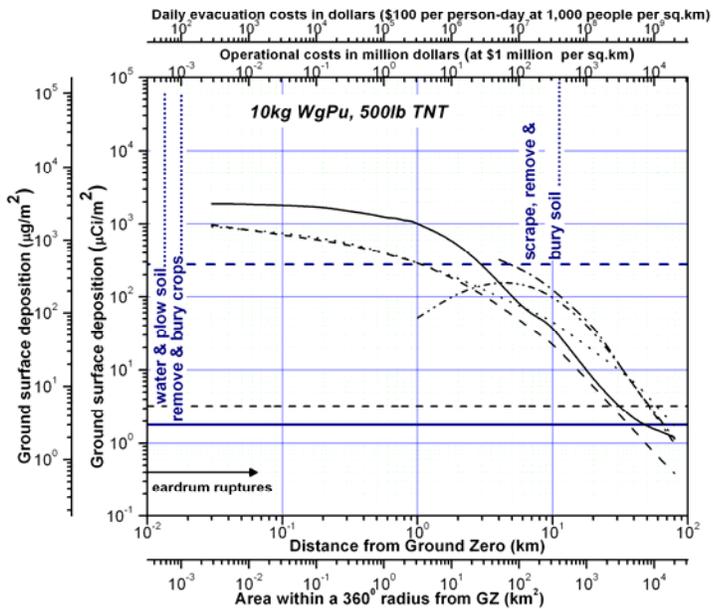

Fig.7. Ground surface deposition with respect to distance from GZ (500 lb TNT)

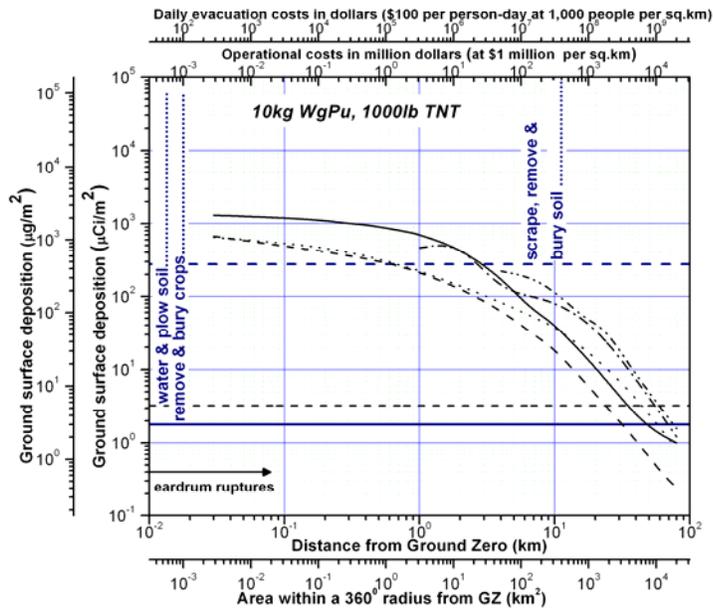

Fig.8. Ground surface deposition with respect to distance from GZ (1000 lb TNT)

verification study of the DIFOUT model has also been performed by the authors of the report and incorporated in the final EIS.

In our study we have normalized the DIFOUT results to a quantity of 10 kg of WgPu and have juxtaposed them with the results of HOTSPOT under the same conditions (unfavorable and median). The results are shown in Fig. 1-8. Four sets of explosive energies of TNT equivalent have been postulated (the same as those adopted by the Pantex Plant reports), namely: 183lb TNT (Fig.1,5), 300lb TNT(Fig.2,6), 500lb TNT(Fig.3,7), 1000lb TNT(Fig.4,8) and for each set of energies we have derived TEDE50 (Fig.1-Fig.4) and surface contamination (Fig.5-Fig.8) charts. In each figure the dash-dot-dotted and dash-dotted curves respresent the DIFOUT model predictions under median and unfavorable weather conditions respectively, defined in the Pantex EIS[22].

Dotted and dashed curves represent the HOTSPOT results for similar weather conditions respectively, while the solid curves are derived applying a worst-case weather scenario[23] to our HOTSPOT simulations. All HOTSPOT simulations assume a fairly realistic distribution of respirable and non respirable material (i.e. the default HOTSPOT activity distribution assuming the plausible scenario where only 20% of the initial WgPu is converted into respirable aerosols with a deposition velocity of 3 mm/s for the respirable fraction and 80 mm/s for the non-respirable one).

In Figs 1-4 we estimate and plot the centerline TEDE50[24] (logarithmic scale) committed to an individual downwind from GZ as a function of distance from GZ for very unfavorable atmospheric

---

[22] Median weather conditions: wind speed=6.75 m/s, Stability category D. Unfavorable weather conditions: wind speed=4.25 m/s, Stability category E

[23] Wind velocity (at 10 meters) u=1m/s, Stability Category F (standard terrain), Mixing layer height=300 m, Dry deposition, recipient height=1.5 m.

[24] TEDE50 is the sum of all dosed committed to the recipient due to inhalation, submersion and ground shine. TEDE50= CEDE50 (inhalation) + EDE (submersion) + EDE (4-days of Ground Shine). TEDE50 is the total effective dose equivalent committed to a recipient within 50 years after his/her exposure.

Page 16 of 30

conditions (assuming a normal breathing rate[25] of $3.3 \times 10^{-4} \, m^3 s^{-1}$. In the same plots the TEDE50 is translated into mass (μg of WgPu, middle left ordinate) and activity (μCi of WgPu, far left ordinate) inhaled[26] by a recipient (at a normal breathing rate). On the right ordinate the committed TEDE50 is translated into the Additional Cancer Risk 50 (ACR50)[27], which can vary between a lower (inner right ordinate) and upper limit (middle right ordinate)[28]. Fig.1-4 also yield an approximate estimate of the daily evacuation costs (top scale) and the number of people that should be evacuated (bottom scale) according to a 360° potential-hazard zone[29] that will be adopted by the authorities (see Sec.6 for details). Alternatively, the bottom scale can be used to estimate the number of people who stand a non-zero probability of receiving a particular TEDE50. These numbers are calculated for a population density of 1,000 people per sq.km, typical of metropolitan city areas and they are linear functions of the population density. For example in areas with ten times larger (smaller) population densities the evacuation costs and the number of the evacuees would be ten times larger (smaller).

Implementing the assumptions analyzed in Sec.6, Figs 1-4 also estimate the minimum health costs per person (far right ordinate).

In Figs 5-8 we plot the ground surface contamination (logarithmic scale) as a function of distance from GZ for the same scenaria[30] described in Fig.1-4. The general notation scheme is the same as in Fig.1-4 while we also indicate three decontamination zones according to the decontamination procedures adopted in the Palomares "broken arrow" event (see table 1). The top scale yields an approximate estimate of the costs involved in the decontamination and evacuation operations according to the analysis in Sect.6. Moreover, under the same assumptions, Figs, 5-8 estimate the evacuation costs due to ground surface contamination and the operational costs entailed in the necessary decontamination procedures. An approximate estimate of the contaminated area (within a 3600 radius from ground zero) is provided at the far bottom scale in Figs 5-8.

---

[25] This is the reference man value for an 8-h workday. For a more elaborate approach age-specific breathing rates might be considered (e.g. $1.17 \times 10^{-4}$ for a child and $4.444 \times 10^{-5}$ for an infant – see Pantex Plant EIS. ibid)

[26] Actually, TEDE50 includes doses from submersion and ground shine but in the case of WgPu they are negligible.

[27] In the framework of the linear risk model the current cancer risk for associated radiation is 0.05 % per rem (ICRP Publication 30, Oxford: Pergamon Press, 1990). Thus if you received a dose of 10 rem, to the whole body, your risk of dying from cancer would increase by 0.5%. Similarly, a dose of 100 rem would increase the risk by 5% while a dose of 300 rem by 15%. This increase will be called in our study additional cancer risk 50 (ACR50). For perspective a whole body CT delivers roughly a CEDE50 of 1 rem to the patient while a chest x-ray exam delivers a dose of 10-20 mrems.
According to the linear risk model, which is the most conservative one, one cancer death will ensue regardless of whether a total dose of 2000 rem is delivered to a single individual or a single dose of 1 rem to a total of 2000 individuals. Thus the cancer risk coefficient CRC50 is:
*CRC50 = 1 Cancer / 2000 person-rem*
In the United States, workers are limited to a whole-body dose of 5 rem per year which, amounts to a CEDE50 of 250 rem (ACR50=12.5%). It is very natural, therefore, to regard the distance from GZ at which an individual receives a CEDE50 of 2000 (250) rem as a reasonably critical (safe) distance since it causes to an individual standing there an ACR50 of 100% (12.5%) in a period of 50 years after exposure.

[28] (Fetter and Von Hippel 1990) assumed 3-12 cancer per mg of WgPu that is inhaled. Our mixture is three year old WgPu (see relevant table). Using the Hotspot rem/Ci ratios we conclude that Fetter and Von Hippel actually assume that it takes a CEDE50 of approximately 2083 rem to 8333rem to cause one cancer.

[29] Even in the case of a radioactive-material release monitored by a grid of meteorological towers and accurate terrain data, evacuation decisions do not typically depend on a predicted plume dogleg to circumvent a community. Rather, the potential concentration as a function of distance is considered, and the evacuation is a remote possibility for the targeted community. In fact, it is not uncommon to evacuate communities based on a 360° potential-hazard zone, effectively eliminating the wind-direction variability problem[29]. This is particularly common at a low wind speeds (for example, <2 m/s) in which wind direction is frequently changing

[30] (Fetter and Von Hippel 1990) didn't use the Gaussian model to derive surface contamination estimates.

Instead of the normal breathing rate, more specific breathing rates can used in Figs 1-4 according to the following table:

| Table 2. Specific breathing rates for humans | | |
|---|---|---|
| **Recepient** | **Activity** | **Breathing rate (m$^3$/s)** |
| Reference man[31] | Resting | $1.25 \times 10^{-4}$ |
|  | Light activity | $3.33 \times 10^{-4}$ |
|  | Heavy Work | $7.1 \times 10^{-4}$ |
|  | Heavy Exercise | $20.0 \times 10^{-4}$ |
| Adult | Averaged over 24 hours | $2.5 \times 10^{-4}$ |
| Teenager | Averaged over 24 hours | $2.5 \times 10^{-4}$ |
| Child | Averaged over 24 hours | $1.17 \times 10^{-4}$ |
| Infant | Averaged over 24 hours | $4.44 \times 10^{-5}$ |

From the above table we can observe that the actual TEDE50 given by Figs 1-4 (and the associate cancer risk ACR50 and health costs) of the exposed population depends also on their activity during the time they are inside the radioactive plume. For example people engaged in heavy work during their contact with the cloud will inhale five as much radioactivity as those resting and therefore their cancer probability will be five times higher compared to those who were resting during their exposure. These scaling factors should be taken into account in all the derived figures and maps of this study, which is a straightforward calculation.

All the results of our simulations shown in the available figures Fig.1-Fig. 8 are actually normalized to 10 kg of WgPu since all the quantities measured on the vertical axes of all figures are linear functions of the initial WgPu mass. If the actual WgPu mass is for example ten times smaller (larger) we simply divide (multiply) the TEDE50 (and all the relevant quantities on the ordinates of Fig.1-4) and the Ground surface densities (and all the relevant quantities in the ordinates of Fig. 5-8) on the vertical axes by ten. Similarly, if the age of the WgPu involved is different (e.g. WgPu in warheads to be decommissioned) then we should multiply the TEDE50 and the activities with the appropriate scaling factors (see appendix).
The arrows in Figs. 1-8 indicate the maximum distance from GZ at which one could experience eardrum ruptures due to the blast wave from the explosion.

**5. The Cylinder Model**
Gaussian dispersion models cannot provide very reliable results at distances very close to the initial cloud. We can provide an approximate estimate of the radiological hazards to individuals standing very close to the initial cloud by a very simple model.
Let us assume that an explosion disperses 10 kg of WgPu. (Church H W 1969) gives the following empirical relationships for the height, $H$ (meters), and radius, $R$ (meters), of clouds formed by high-explosive detonations:

$$H = 92.6 W^{0.25} \, m \qquad R = 4.7 W^{0.375} \, m \qquad (0.1)$$

where $W$ is the explosive yield in kilograms of TNT equivalent.

---

[31] ICRP 1974

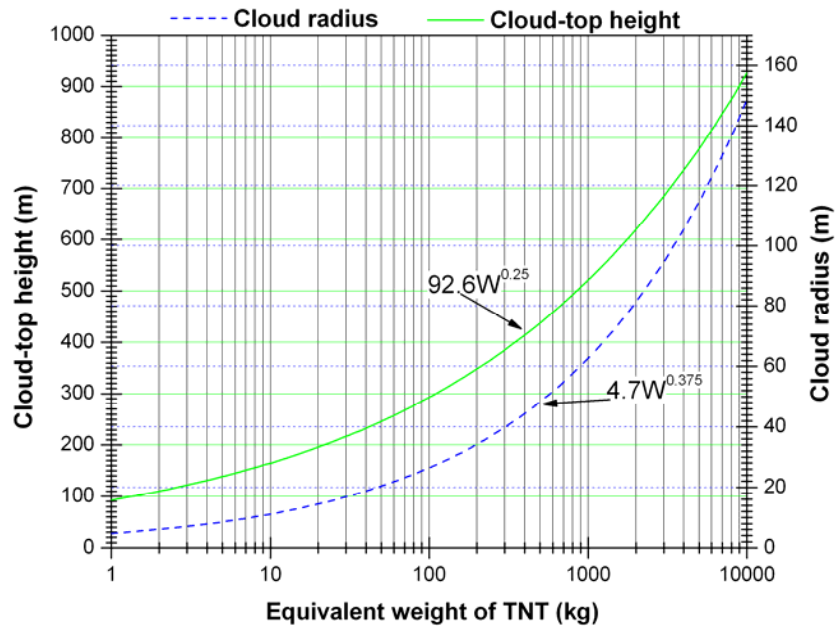

Fig.9. The cloud-top height (solid curve, left-hand-side ordinate) and the cloud radius (dashed curve, right-hand-side ordinate) with respect to the TNT equivalent energy of an explosion

Experiments have shown[32] that approximately 5% of the radioactivity in the cloud is initially found between the ground and T/4, where T is the cloud-top height; 30 percent between T/4 and T/2; 40 percent between T/2 and 3T/4; and 25 percent between 3T/4 and T.

Thus, the average concentration of aerosol near the ground, $\chi$ (milligrams per cubic meter), is roughly

$$\chi_{(mg/m^3)} = \frac{0.05 f_i M_{(kg)}}{\pi R^2 (H/4)} = \frac{68 f_i M_{(kg)}}{W_{(lb)}} = \frac{31 f_i M_{(kg)}}{W_{(kg)}} \quad (0.2)$$

where $M$ is the mass of the penetrator (kilograms) and $f_i$ is the fraction converted to respirable aerosol.

Now consider a person standing in the open, directly downwind from the center of the cloud (assumed to retain its density, shape, and dimensions during its transition, which is a conservative scenario). This person would be immersed in the cloud for a maximum time $t \cong 2R/u$, where u is the wind speed (meters per second). Assuming that the density distribution of the aerosol remains constant[33], the maximum total amount of aerosol inhaled, $I$ (milligrams), during this time would be

$$I_{(mg)} = \chi_{(mg/m^3)} t_{(s)} b_{(m^3/s)} = \frac{2Rb\chi_{(mg)}}{u} = \frac{291 fbM_{(kg)}}{uW_{(kg)}^{0.625}} \quad (0.3)$$

---

[32] Pantex Plant EIS ibid.

[33] This assumption practically means that the person is standing very close to the initial cloud and that the cloud radius is very small.

Page 19 of 30

where b is the assumed breathing rate ($3.3 \cdot 10^{-4} \, m^3/s$) for an adult male performing light activity. Higher values of $W$ result in lower inhaled doses because the energy release disperses and dilutes the aerosol.

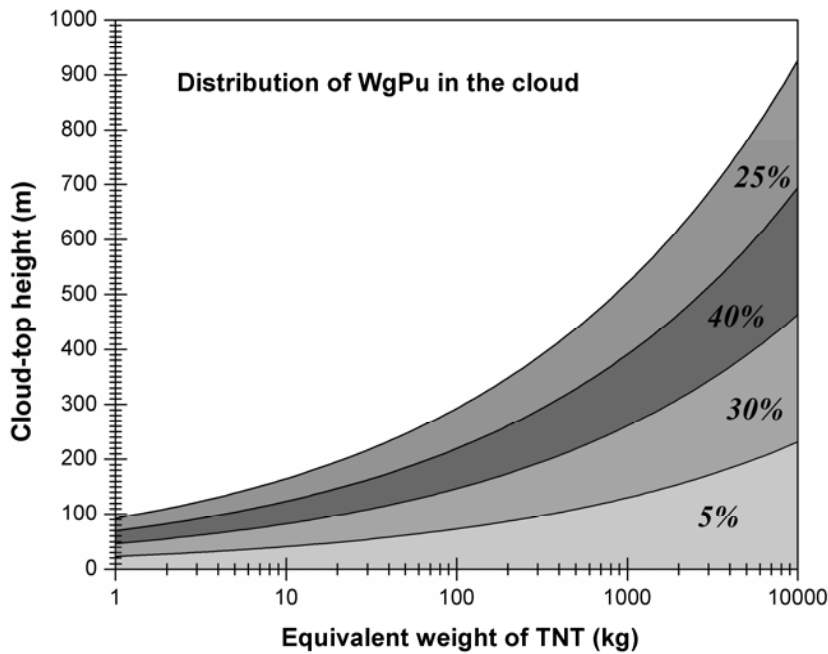

Fig.10. The percentage of the radioactivity in the cloud with respect to the TNT equivalent energy of an explosion. The first (bottom) shaded area (5%) represents the cloud layer between the ground and T/4, the second shaded area above the ground one represents the cloud layer between T/4 and T/2 (and so forth, see text)

Figure 11 depicts the TEDE50 (and the related values i.e. activity, mass etc, see previous figures) committed to an individual at very close distances after the explosive dispersion of 10 kg WgPu (all airborne and respirable) with respect to the TNT equivalent energy (according to the cylinder model). Moreover, the radii and the maximum heights of the cloud as a function of the TNT equivalent energy yield are shown on the bottom and top scales, respectively. The inhaled quantity of WgPu and the associated risk scales linearly with the initial quantity of WgPu dispersed. The far bottom scale indicates the cloud of the radius which can be considered the effective distance at which the individual should be standing in order to receive the TEDE50 given in the plot. For example the explosive dispersion of 10 kg of WgPu dispersed by an explosion of 100 kg TNT equivalent would generate a cloud of an (approximate) radius of 28 m. A person standing still at such a distance during the passage of the (rigid) cylindrical cloud would run a risk of receiving a TEDE50 larger than 1000 rem , close the threshold limit of 2000 rems.

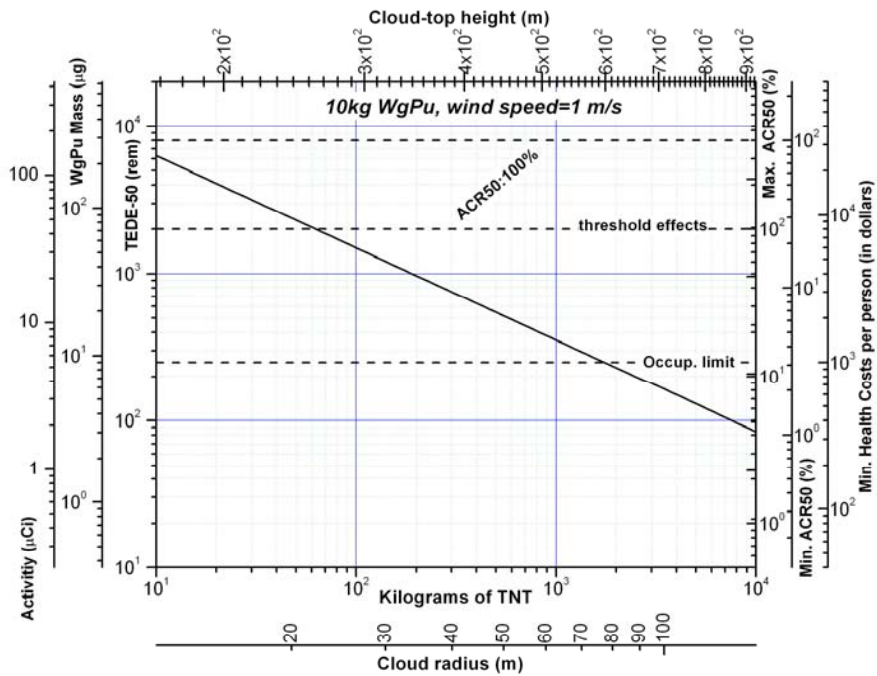

Fig.11. The TEDE50 (and the related values) committed to an individual at very close distances after the explosive dispersion of 10 kg WgPu (all airborne and respirable) with respect to the TNT equivalent energy (according to the cylinder model).

**6. Health, Decontamination and evacuation costs after a "broken arrow".**

*Health Costs.*

According to the National Cancer Institute[34] researchers compiled the treatment costs for 932 adult cancer patients who were enrolled between Oct. 1, 1998, and December 31, 1999, in trials sponsored by the National Cancer Institute (NCI). They then compared these costs with those for 696 cancer patients who were treated outside of clinical trials. For each patient, treatment costs were calculated for a period averaging 2.5 years from the date of the individual's cancer diagnosis. Included were doctor visits, hospital stays, diagnostic tests and procedures, and all drugs given in a doctor's office, hospital outpatient department, or other treatment settings. Patients' out-of-pocket costs were included as well as costs reimbursed by third-party payers. The researchers concluded that treatment costs for trial participants were 6.5 percent higher than they would have been if these patients had not enrolled in a trial: $35,418 for trial participants and $33,248 for nonparticipants (www.cancer.gov) (Goldman et al 2003.). Therefore in the framework of the linear risk model we can make the plausible assumption that the average treatment cost per cancer is approximately $34000 (2003 FY dollars). Since we have one cancer per 2000 man-rem-8000 man-rem we actually have a treatment cost of $4-$17 (2003 FY dollars) per man-rem received by the exposed population.

*Evacuation Costs.*

---

[34] www.cancer.gov

In contrast to popular beliefs actual evacuations from disaster areas other than nuclear have been carried out consistently in orderly fashion without drills. The motivation of the public to participate in drill evacuations appears to be very low. Common sense indicates that evacuation costs can be roughly approximated by the sum of the costs incurred for sheltering, food supplies and first aid. An older investigation (Fischer-Colbrie, E 1979) of the expected cost and effectiveness of exercise evacuations from the vicinity of nuclear power plants led to the following results and conclusions: the estimated total cost per person and day varies from $17.34 to $78.20 (1979 FY dollars). Thus the indicative amount of 100$ per day-person used in this study is obviously a fairly reasonable assumption.

*Decontamination and Operational Costs.*

These are the most formidable costs to evaluate. In this study there has been an inherent assumption that the high explosive of the warhead has exploded on the ground. The Palomares and the Greenland experience have indicated that Ground Zero can actually be on top of the hills, in the depths of the sea or under a thick layer of ice and snow. Operation costs should also take into account the claims and demands made by the population in the vicinity of GZ and their government regarding the effects of such an accident. In view of the difficulty in devising a model which would provide a reasonable cost per contaminated area we have simply treated that cost as a free parameter (one million dollars per square kilometer).

**7. Postulating a Minot Base "Broken Arrow" event: Analysis & Conclusions**

**7.1 Dose effects**

The analysis in the appendix indicate that the six W80 nuclear warheads reportedly mounted on the six ACMs at the USAF Minot Base could not possibly contain more than 36 kg of WgPu (six kg of WgPu each) with a more realistic approximate estimate of 18 kg WgPu (3kg of WgPu each). Therefore all the results in Figs 1-8 should be scaled appropriately for a relevant postulated "broken arrow". This can be accomplished by multiplying all the readings on the ordinates by 1.8 (realistic estimate of WgPu) or 3.6 (maximum possible quantity of WgPu), which then in turn modifies the relevant readings on the top and bottom scales (costs, areas, number of people, etc.).

According to Fig.1-4, even under the most unfavorable weather conditions (absolutely worst-case scenario adopted by the HOTSPOT model, i.e. solid curve), the simultaneous explosive dispersion of all the WgPu contained in the primary devices of the six thermonuclear warheads (36 kg of WgPu) would generate a radioactive plume that could note pose any serious immediate threat (due to inhalation, submersion or ground shine) at distances larger than one kilometer from GZ. This result is consistent with the estimates of (Fetter S and Hippel F V 1990). However, according to such a worst-case scenario of HOTSPOT, all the people at distances shorter than one kilometer from GZ would be at a non-negligible risk of inhaling WgPu aerosols delivering a total TEDE50 of (approximately) 360 rem per person (or larger), thus increasing the individual cancer risk by (at least) an amount of 5% -20% (minimum and maximum values read on the right ordinates). The minimum health costs per person who receives a TEDE50 of 360 rem would amount to (approximately) $1,500, or higher. All populated areas within a radius of one kilometer from GZ should be evacuated and the population should undergo medical tests to ensure that they have not inhaled any WgPu aerosols. In large metropolitan cities (1,000 people/sq.km) this amounts to 3,140 people and a total daily evacuation cost of $314,000.

Using the more realistic assumption of a total quantity of 18 kg of WgPu the TEDE50 values (and all the relevant values on the left and right ordinates) would be approximately doubled but this would not alter the general picture since, given accurate input assumptions, the standard deviation of the TEDE50 dose values as calculated in Hotspot is approximately a factor of 5 (HOTSPOT), while some authors report a factor of 3 (Cember, 1985). Therefore, 68% of the time (i.e., the percentage of observations within 1 standard deviation, assuming a Gaussian distribution), the calculated dose values will be within a factor of 5. This level of accuracy is more than acceptable for emergency response and planning.

The DIFOUT model predictions are even more optimistic excluding any inhalation hazard under any weather conditions at distances larger than one kilometer. Unfortunately, the radiological risk estimates derived by the more realistic models of DIFOUT disregard all areas within a radius of one kilometer from GZ. However, comparing the HOTSPOT predictions with the available DIFOUT ones we observe that the former consistently overestimate the latter so that at some distances the HOTSPOT TEDE50 (and all the relevant quantities on the ordinates) may actually be up to an order of magnitude larger than the DIFOUT ones. It is therefore reasonable to adopt the HOTSPOT TEDE50 predictions as the absolutely worst-case scenaria in our study bearing in mind that the actual TEDE50 (and all the relevant quantities on the ordinates) may be up to an order of magnitude smaller.

Both HOTSPOT and the cylinder model predict increased cancer risk at distances very close to GZ (less than 100 m), where the health effects of the blast will also become apparent (e.g eardrum rupture). Adopting the plausible assumption that only 20% of the maximum amount of 36 kg of WgPu would become respirable then Figure 11 (cylinder model) approximates the effects of the respirable cloud on humans close to GZ. It is obvious that an explosion less energetic than a few hundreds of kilograms of TNT equivalent can certainly commit TEDE50 doses (to individuals close to GZ) well above the occupational limit. Especially at very short distances (less than 40 m) from GZ, the cylinder model predicts that all people run a risk of inhaling lethal doses of WgPu aerosols which could initially cause acute health effects and eventually cancer with certainty.

Reading Figs 1-4 one might draw a multitude of conclusions about the health risks that would have resulted if the Minot Base "Bent Spear" had actually been a worst-case "Broken Arrow", such as:

a) there is a non-zero probability[35] that an individual standing at a distance of 1 km downwind from GZ will receive TEDE50 above 250 rem (occupational limit, approximately 12% ACR50) in the next 50 years of his life (due to inhalation of the radioactive cloud, submersion in it and ground shine). However, the more realistic predictions of DIFOUT indicate a TEDE50 which are at least an order of magnitude smaller
b) it is rather impossible for a person standing at distances downwind from GZ larger than 0.2 km to receive a TEDE50 of 2000 rems, which would both cause acute radiation sickness and increase the person's probability of developing cancer in the next 50 years by 100%!
c) if the authorities do decide to evacuate all people who might receive a TEDE50 of 250 rem (occupational limit) in a city with an average population density of 1,000 people/sq.km then such an evacuation area should extend to a radius of one kilometer from GZ. Ground contamination hazards could extend the evacuation radius even further (see subsection 7.2)

(Fetter & Von Hippel 1990) have already applied the Gaussian model in warhead accident simulations assuming various weather conditions and an explosive energy of 40 kg TNT, which is smaller than those adopted in the Pantex EIS and in the present study. They actually simulated the explosive release of 10kg of respirable WgPu while in our HOTSPOT simulations only 20% of the 10 kg WgPu is assumed to be respirable, which is the default and most probable distribution in WgPu explosion. We were able to reproduce their results running HOTSPOT under similar weather conditions thus verifying that our results compare well with those of other authors.

## 7.2 Ground Contamination effects

Assuming the simultaneous explosion of the (maximum) amount of 36 kg of WgPu, Fig.5-8 show that under the realistic simulations of DIFOUT there is a non-zero probability that some areas within a radius of five kilometers from GZ can be so heavily contaminated with WgPu that their soil should be scraped, removed and buried in a safe location (e.g the **Savannah River Site**). Such a decontamination procedure, which should invariably be preceded by radiation detection procedures, would definitely force the authorities to evacuate the population within at least a radius of five kilometer from GZ. Using the more realistic assumption of a total quantity of 18 kg of WgPu the results are almost the same due to the shape of the DIFOUT curves (see relevant figures).

---

[35] In fact that probability becomes certainty if the assumed meteorological conditions are true and the wind blows constantly and directly towards an individual who never leaves his/her position throughout the entire plume passage.

Therefore, even if the WgPu inhalation hazard cannot extend to such large distances, ground contamination should definitely stretch evacuation distances to five kilometer from GZ. At distances larger than ten kilometers from GZ, there is a non-zero probability to find contaminated soil which should be watered and plowed and crops that should be removed and buried. No agricultural products from such areas should reach the market unless they have been screened for radiological contamination. Large scale emergency procedures at a metropolitan city (1,000 people per sq.km.) would entail the evacuation of citizens living in areas of seventy five square kilometers, that is 75,000 people. A daily total evacuation cost of $7,500,000 should be anticipated by the authorities while the decontamination operational costs would be of the order of $75,000,000 (assuming arbitrarily an operational cost of one million per sq.km which should be scaled accordingly).

**Acknowledgements**
This work was conducted by the author at the Center for International and Security Studies at Maryland (CISSM) during his sabbatical leave from the Hellenic Army Academy. It was Prof. Steve Fetter's idea that the author should focus on the effects of a postulated "broken arrow" that might ensue from the Minot USAF Base "bent spear" event. The author is grateful to Prof. Fetter for his encouragement, advice, and guidance as well as to the Director of CISSM Prof. John D. Steinbruner for his kind hospitality and support.

# Appendix

**A. Fissile material and weapon model.**

A.1 Nuclear Weapons design

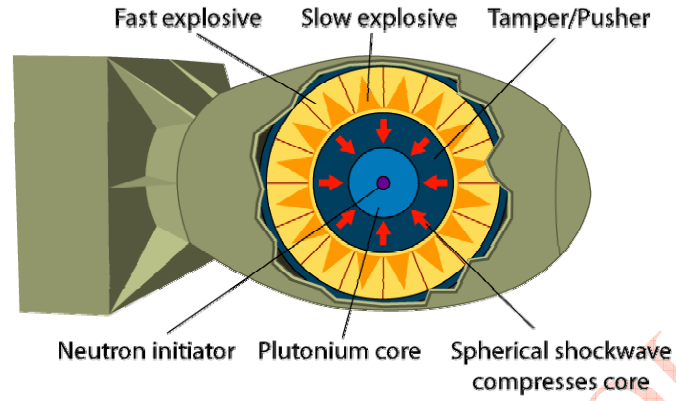

*Naïve representation of the first implosion fission weapon used in combat (Fat Man). The same physics principles (with many improvements of course) are used in the primary device (trigger) of modern thermonuclear weapons (www.wikipedia.org).*

A.1.1. **Implosion assembly.** In order to assess the effects and the risk of a "broken arrow" it is imperative that some general facts about the nuclear weapon involved in the accident are known. Modern thermonuclear weapons design is of course classified, however, in a worst-case scenario attempting to derive order-of-magnitude estimates of the radiological effects all we need to know is the type and the (maximum possible) quantity of the nuclear materials involved in the accident as well as the energy yield of the chemical explosion which will disperse the nuclear material in the vicinity of the accident. According to the open literature modern thermonuclear weapons (hydrogen bombs) consist of two major components: the primary and the secondary device. The primary device, which is the trigger of the fusion weapon, is believed to be an implosion-type fission explosive. In an implosion assembly a subcritical spherical, or sometimes cylindrical, mass of weapon-grade plutonium is compressed by using specially designed lenses of high explosives. Implosion works by initiating the detonation of the explosives on their outer surface, so that the detonation wave moves inward. Careful design allows the creation of a smooth, symmetrical implosion shock wave. This shock wave is transmitted to the fissionable core and compresses it, raising the density to the point of supercriticality. Implosion can be used to compress either solid cores of fissionable material, or hollow cores in which the fissionable material forms a shell. The "gadget" (the fist plutonium bomb tested in Alamogordo ) as well as in Fat Man dropped on Nagasaki were both implosion assembly weapons. An implosion fission weapon may require a source which can produce a precisely timed burst of neutrons to initiate the chain reaction in the plutonium core. The type of neutron initiator used in early implosion devices utilized the emission of neutrons caused by bombardment of Berylium-9 or some other light element by alpha particles. This requires a strong source of alpha particles, something of the order of 10 curies of Polonium-210 or a similarly active alpha emitter (Military Critical Technologies DoD 1998). This isotope of polonium has a half life of almost 140 days, and a neutron initiator using this material needs to have the polonium replaced frequently. To supply the initiation pulse of neutrons at the right time, the polonium and the beryllium need to be kept apart until the appropriate moment and then thoroughly and rapidly mixed. Modern thermonuclear warheads do not use Po-210.

Modern nuclear warheads are one-point safe which means that when the high explosive lenses of the warhead are detonated at any single point, the probability of producing a nuclear yield exceeding 4 pounds of TNT equivalent is less than one in a million. Since $2.5\times10^{17}$ fissions yield an energy of 4 pounds of TNT a nuclear warhead accident would also produce small amounts of fission products (notably iodine, strontium, and cesium) from a fission yield not exceeding $2.5\times10^{17}$ fissions per warhead. Organ doses acquired by inhalation of these fission products are negligible fractions (less than 1 ‰) of the plutonium dose (Elder *et al* 1982) and will not be considered in this study

**Weapon Grade Plutonium**

The fission primary of a modern thermonuclear weapon use weapon grade plutonium.
The critical mass of a bare sphere[36] of weapon grade plutonium is about 16.28 kilograms. Surrounding the WgPu sphere with a metallic spherical shell (tamper) to reflect the escaping neutrons back into the fissile material the critical mass would be reduced according to the following table (Paxton 1964):

| Table 3. Plutonium Criticality | | | | | | |
|---|---|---|---|---|---|---|
| Plutonium Core[37] | | Reflector (Tamper)[38] | | | | Critical Mass |
| w/o Pu-240 | Density | Material | Shape | Thickness | Density | |
| 4.5 | 15.66 g/cm$^3$ | None | - | - | - | 16.28 kg |
| 4.8 | 15.36 g/cm$^3$ | U(N) | Sphere | 7.72 in. | 19.0 g/cm$^3$ | 5.91 kg |
| 4.9 | 15.62 g/cm$^3$ | Be | Sphere | 1.45 in | 1.83 g/cm$^3$ | 8.39 kg |
| 4.9 | 15.74 g/cm$^3$ | Al | Sphere | 3.12 in | 2.82 g/cm$^3$ | 11.04 kg |

Fat Man (the bomb dropped on Nagasaki) used 6.1 kilograms of plutonium (Serber 1992). Modern weapons undoubtedly make more efficient use of plutonium by using reflectors according to table 3. Therefore assuming that the fission primary device of a thermonuclear weapon contains 6 kilograms of WgPu is definitely consistent with a worst-case scenario although more realiable estimates (Cochran T B 1998) for a high-tech fission primary device indicate that the Fat Man yield can be achieved by a high-tech implosion-type fission weapon containing no more than 3 kilograms of WgPu.

All figures have been derived for a mixture of WgPu @ 3 years. If an older mixture is used the mass inhaled by a recipient downwind is actually the same, however the TEDE50 is different (and so is the associated alpha and total activity inhaled). Thus, for WgPu @ 25 years (typical for warheads before decommissioning) the TEDE50 at a particular distance downwind is approximately 10% larger than the TEDE50 of WgPu @ 3 years, which means that all the TEDE50 readings of Fig.1 should be increased[39] by 10% when it comes to WgPu @ 25 years.

---

[36] The bare critical mass is not the mass one would need to construct a device since by the use of a neutron reflector (tamper: a spherical metallic shell surrounding the fissionable material) the critical mass can be reduced by at least a factor of two.

[37] w/o weight percent

[38] U(N): Natural Uranium, Be:Berelium, Al:Aluminum

[39] The WgPu default explosion scenario of the HOTSPOT code (S. Homman, private communication) ignores the 10% beta component (Pu-241) and uses the total alpha activity (0.081 Ci/g @ 3 years) and the dose conversion for Pu-239. We have been able to verify that the default WgPu actually disregards all Pu-241 and assumes that the WgPu consists exclusively of Pu-239. Such an assumption underestimates the TEDE50 committed by WgPu by (approximately) 10% , which is what you have estimated. In the author's opinion, HOTSPOT users should be warned that the integrated air concentration [(Ci-s)/m3] and ground contamination calculated by the default WgPu explosion (under the above assumptions) only includes the alpha activity

| Table 4. Plutonium data | | | | | | | |
|---|---|---|---|---|---|---|---|
| **One kg of (3-year / 25-year)-old WgPu[40]** | | | | | | | |
| *Isotopes* | *Half-Life (years)* | *Sp. Activity (α,β) (Ci/g)* | *rem/Ci* | *rem/g* | *Mass (g)* | *Activity (Ci)* | *CEDE50 (rem)* |
| Pu-238 | 87.74 | **α:**17.12 | $2.88 \times 10^8$ | $4.93 \times 10^9$ | 0.391/0.328 | 6.694/5.615 | $1.93 \times 10^9 / 1.62 \times 10^9$ |
| Pu-239 | 24065 | **α:**0.06205 | $3.08 \times 10^8$ | $1.91 \times 10^7$ | 933.32/932.72 | 57.892/57.892 | $17.8 \times 10^9 / 17.8 \times 10^9$ |
| Pu-240 | 6537 | **α:**0.227 | $3.08 \times 10^8$ | $6.99 \times 10^7$ | 59.98/59.84 | 13.620/13.574 | $4.19 \times 10^9 / 4.19 \times 10^9$ |
| Pu-241 | 14.35 | **β:**102.3 | $4.96 \times 10^6$ | $5.07 \times 10^8$ | 5.018/1.734 | 513.546/176.979 | $2.54 \times 10^9 / 8.79 \times 10^8$ |
| Pu-242 | 376300 | **α:**0.003938 | $2.93 \times 10^8$ | $1.14 \times 10^6$ | 0.40/0.40 | 0.0015/0.0015 | $4.56 \times 10^5 / 4.56 \times 10^5$ |
| Am-241 | 432.2 | **α:**3.428 | $4.44 \times 10^8$ | $1.52 \times 10^9$ | 0.78/3.97 | 2.670/13.609 | $1.19 \times 10^9 / 6.03 \times 10^9$ |
| | | | | | | | |
| *T* | *O* | *T* | *A* | *L* | 1000/1000 | 595/269 | $27.7 \times 10^9 / 30.5 \times 10^9$ |
| | | | | | | | |
| **3-year-old WgPu** | | | | **25-year-old WgPu** | | | |
| $2.77 \times 10^7$ rem/g | | | | $3.05 \times 10^7$ rem/g | | | |
| 0.595 Ci/g | | | | 0.269 Ci/g | | | |
| $4.63 \times 10^7$ rem/Ci | | | | $1.14 \times 10^8$ rem/Ci | | | |
| (WgPu α-activity)=0.033×(α-activity of Am-241) | | | | (WgPu α-activity)=0.150×(α-activity of Am-241) | | | |
| | | | | | | | |
| Conversion factors[41] | | | | | | | |
| (Total activity of WgPu @ 3 years) = 2.22 ×(Total activity of WgPu @ 25 years) | | | | | | | |
| (TEDE50 of WgPu @ 3 years) = 0.91×(TEDE50 of WgPu @ 25 years) | | | | | | | |

On the other hand Fig.2 depicts total activity ground surface concentrations for a mixture of WgPu @ 3 years. When it comes to a mixture of WgPu @ 25 years all activity (and mass) concentration should be approximately halved (divided by a factor of 2.22). According to HOTSPOT, if D is the calculated radiation dose, then 50% of the time the true dose should lie between D/3 and 3D; and 80% of the time between D/8 and 8D. Therefore, taking into account the Gaussian model errors we can plausibly assume that all figures can be used for WgPu accidents whose age ranges from zero to several decades[42].

**Tritium**
Soon after the design and testing of the first fission weapons it was realized that the extremely high temperature attained during a fission explosion can be used to initiate thermonuclear fusion of light

---

and disregards the beta activity of Pu-241. Using a realistic mixture of WgPu yields much different (larger) air concentrations and ground contamination since it also takes into account the beta activity of Pu-241.

[40] The default WgPu explosion scenario of HOTSPOT makes the following assumption (S.Homann, private communication): It ignores the beta activity of Pu-241 and only takes into account the alpha activity of WgPu @ 3 years (0.081Ci/g). Then HOTSPOT uses the conversion factors of Pu-239 in calculating the TEDE50.

[41] The CEDE50 dose is (90%) due to the activity of alpha emitters (Pu-238, Pu-239, Pu-240, Pu-242, and Am-241) and (10%) due to the activity of the beta emitter Pu-241. Pu-241 beta decays into Am-241 thus reducing the total activity of the mixture buy simultaneously increasing the total TEDE50 of the mixture (see table).

[42] For example, in the case of WgPu @ 25 years Fig.1 underestimates the relevant TEDE50 by a negligible percentage of 10% while Fig.2 overestimates the total activity by 220% (which actually yields more conservative estimates of ground surface concentrations)

nuclei[43]. In fact it is well known that the rate of thermonuclear energy production rate is an increasing function of density and temperature (Clayton 1984). This fact led weapons designers to vent a small amount of deuterium[44] and/or tritium into the plutonium pit of an implosion assembly. The temperature attained during a fission explosion is larger than that existing in the center of the sun and thus it was enough to ignite the thermonuclear fusion of deuterium and/or tritium just as it happens in the hydrogen-burning zone of ordinary stars[45]. In a typical modern boosted fission primary, a few grams of deuterium-trimium gas are injected into the center of a hallow core of plutonium immediately prior to the detonation of the surrounding chemical explosives. The fission primary of the W80 Mod-1 thermonuclear warhead is reportedly a variable-yield (or dial-a-yield)[46] boosted fission weapon. According to open-source information the W88 warhead with fresh tritium inside its pit may explode with a yield of 475 kilotons, but with no tritium inside the core it might explode with the force of just 20 kilotons. In the same way the W80 Mod-1 warhead involved in our study can yield either 7KT or 150 KT of TNT.

Unlike deuterium which is a non-radioactive isotope, tritium is extremely radioactive and beta decays to Helium-3 (mean beta particle energy 5.7 keV; decay energy 18.6 keV). Due to the low energy of its beta decay tritium does not pose an *external* radiation hazard because the charged decay products have a very small mean free path in water or a similar shield. However, if *tritiated water vapor* is inhaled or absorbed through the skin it can pose an *internal* radiation hazard. Therefore, it is reasonable to believe that in a broken arrow we could be also facing a tritium release risk. It is also common sense that in thermonuclear weapons the primary device is only the trigger and therefore it is not expected to yield energies larger than a few tens of kilotons TNT equivalent. Since a few grams of deuterium and/or tritium can enhance a primitive Fat Man device by several times we believe that the primary device in a modern fission-fusion-fission weapon cannot contain more than five grams of tritium (although a more realistic prediction would be of the order of two or three grams-see Bucharin O 2001).

**Refereneces**

Bucharin O 2001, "Downsizing Russia's nuclear warhead production infrastructure", The non-proliferation review Spring 2001

Cember, H. (1985), Introduction To Health Physics, 2nd. ed., p. 350 (Pergamon Press, Oxford).

Church H W 1969 Cloud Rise from High-Explosives Detonations, Sandia Laboratories, TID-4500, p. 14 (53rd ed., UC-41, Health and Safety, SC-RR-68-903).

---

[43] Fusion boosting is a technique for increasing the efficiency of a small fission bomb by introducing a modest amount of deuterium-tritium mixture inside the fission core. As the fission chain reaction proceeds and the core temperature rises at some point the fusion reaction begins to occur at a significant rate. This reaction injects fusion neutrons into the core, causing the neutron population to rise faster than it would from fission alone.

[44] The hydrogen atom consists of a nucleus and an orbiting electron. There are various isotopes of hydrogen such as natural hydrogen which has only a proton in its nucleus, deuterium (one proton+one neutron) which exists in heavy water, tritium (one proton+2 neutrons).

[45] Fusion reactions, also called thermonuclear reactions, are reactions between the nuclei of certain isotopes of light elements. If the nuclei collide with sufficient energy (provided by heat in a bomb, or by a particle accelerator in the laboratory) then there is a significant chance that they will merge to form one or more new nuclei with the release of energy. Different nuclei combinations have different inherent probabilities of reacting in a collision at a particular temperature. The rates of all fusion reactions are affected by both temperature and density. The hotter and denser the fusion fuel, the faster the fusion "burn". See D.D.Clayton , ibid.

[46] Variable yield, or Dial-a-yield, an option available on most modern nuclear weapons, allows the operator to specify a weapon's yield, or explosive power, allowing a single design to be used in different situations. Variable yield technology has existed since at least the 1960s. Examples of variable yield weapons include the B61, B83, W80 and W85 warheads. (wikipedia.org)